\documentclass[11pt,twoside]{article}
\usepackage{asp2010}

\resetcounters

\begin{document}

\title{SPICA spectroscopic cosmological surveys to unravel galaxy evolution}
\author{Luigi~Spinoglio$^1$, Kalliopi~Dasyra$^2$, Alberto~Franceschini$^3$, Carlotta~Gruppioni$^4$, Matt~Malkan$^5$ and Roberto~Maiolino$^6$
\affil{$^1$IAPS - INAF, Via Fosso del Cavaliere 100, I-00133 Roma, Italy}
\affil{$^2$Obs. de Paris, LERMA, 61 Avenue de l'Observatoire, F-75014, Paris, France}
\affil{$^3$Dip. di Astron., Univ. di Padova, v. dell'Osservatorio 5, 35122 Padova, Italy} 
\affil{$^4$Osservatorio Astron. di Bologna-INAF, Via Ranzani 1, 40127 Bologna, Italy}
\affil{$^5$Dept. of Physics \& Astronomy, UCLA, Los Angeles, CA 90095-1547, USA}
\affil{$^6$Cavendish Lab., Cambridge Univ., 19 Thomson Av., Cambridge CB3 0HE, UK}}

\begin{abstract}
The main energy-generating mechanisms in galaxies are black hole (BH) accretion and star formation (SF) and the interplay of these processes is driving the evolution of galaxies. 
MIR/FIR spectroscopy are able to distinguish between BH accretion and SF, as it was shown in the past by infrared spectroscopy from the space by the {\it Infrared Space Observatory} and {\it Spitzer}. 
{\it Spitzer} and {\it Herschel} spectroscopy together can trace the AGN and the SF components in galaxies, with extinction free lines, almost only in the local Universe, except for a few distant objects.
One of the major goals of the study of galaxy evolution is to understand the history of the luminosity source of galaxies along cosmic time. This goal can be achieved with far-IR  spectroscopic cosmological surveys.
SPICA in combination with ground based large single dish submillimeter telescopes, such as CCAT, will offer a unique opportunity to do this.
We use galaxy evolution models linked to the observed MIR-FIR counts (including {\it Herschel}) to predict the number of sources and their IR lines fluxes, as derived from observations of local galaxies.
A shallow survey in an area of 0.5 square degrees, with a typical integration time of 1 hour per pointing, will be able to detect thousands of galaxies in at least three emission lines, using SAFARI, 
the far-IR spectrometer onboard of SPICA.
\end{abstract}

\section{Introduction}

One of the major goals of the cosmological studies of galaxy evolution is to understand the full cosmic history of energy generation by stars (through the fusion process) and black holes (through accretion of matter). This history cannot only be retrieved by high luminosity objects, i.e. the quasars, but mainly by low to intermediate luminosity galaxies, such as those objects corresponding to the Seyfert galaxies in the local Universe, which dominate at the ÔkneeÓ of the Luminosity Function. 
The importance of measuring these energy production rates lies also in the fact that these provide  a measure of the built up of the mass of the central black hole, on one side, and of galactic stars, on the other side and must--ultimately--be consistent. This will lead us to understand the inter-relation of quasar activity and star formation, and ultimately the key processes responsible for shaping the mass and luminosity functions of galaxies.

Optical continuum measurements alone are completely inadequate to obtain these data and even optical spectroscopy on a massive scale cannot yield definitive answers because dust reddening may block our view at short wavelengths. What is needed therefore is spectroscopy at longer rest wavelengths to uncover how much of this emission is partly or heavily extinguished.

\section{Why we need SPICA?}

\subsection{Comparing different techniques and regimes for separating AGN and SF}

No single criteria can be used to distinguish AGN and SF, but there are limits and potentialities of different observational techniques:\\
-- UV/Optical/NIR observations are able to measure galaxy morphology and spectra, however they seriously suffer from dust obscuration.\\
-- X-ray observations are good tracers of AGN, however only weak X-ray emission can be detected from star formation and, even more importantly, heavily-obscured AGN (Compton-thick) are completely lost.\\
-- Radio observations (with planned facilities like EVLA, SKA) can detect AGN and SF to large z and can see through gas and dust, they can measure morphology and spectral energy distributions (SED), detect polarization and variability, however not always redshifts can be measured. At its highest frequencies, SKA could be able to measure redshifted molecular lines in the ISM of galaxies.\\
-- mm/submm observations (e.g. ALMA, CCAT) will provide spectra from SF (redshifted CO, [CII], etc.), however we need to find AGN tracers at the longest FIR wavelengths. 
One candidate is CO: spectral line energy distributions (SLED) are in fact different from PDR (SF) and XDR (AGNs). 
Another candidate can be OH at 119 $\mu$m (that at redshift of z $\sim$ 2 gets into the 350 $\mu$m atmospheric window) that can measure high velocity AGN driven outflows \citep[see, e.g.,][]{fischer10, spoon13}.\\
-- Rest-frame MIR/FIR imaging spectroscopy can provide a complete view of galaxy evolution by measuring the role of BH and SF because it can (provided that large field of view and high sensitivity can be reached) trace simultaneously both SF and AGN, measure redshifts  and see through large amounts of dust. It seems therefore to be the most promising technique. \\

\subsection{The power of infrared spectroscopy}

\begin{figure}[!ht]
\begin{center}
   \plottwo{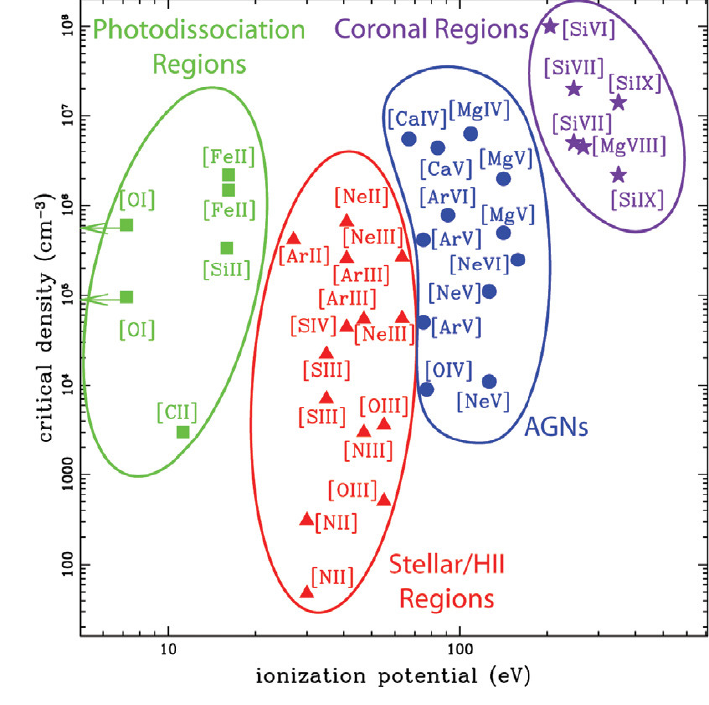}{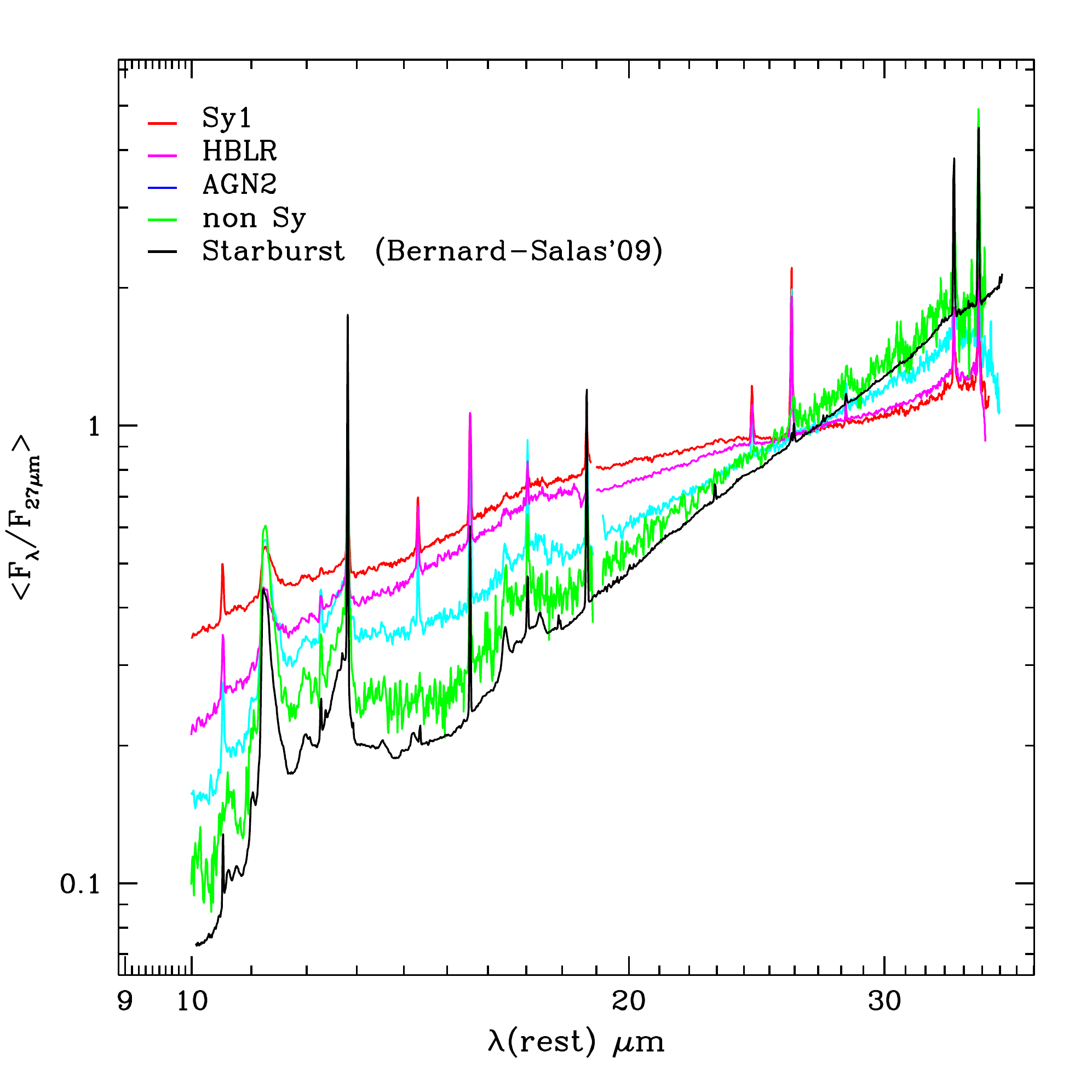}
\end{center}
\caption{
\textit{Left: \textbf{a}}\, Critical density for collisional de-excitation vs. ionization potential of IR fine-structure lines, showing the diagnostic power of the infrared fine-structure lines to trace different astrophysical conditions: from photodissociation regions, to stellar/HII regions, to AGN environments and high excitation coronal line regions \citep{sm92}.
\textit{ Right: \textbf{b}}\, Mid-IR spectra of Seyfert galaxies in the local Universe normalized at 27$\mu$m, showing a sequence with decreasing level of non-thermal activity, from Seyfert type 1's through Hidden Broad Line Region (HBLR) galaxies and type 2 AGN to low luminosity AGN (non-Seyfert's), compared to those of starburst galaxies \citep{b-s09}. 
The bright emission features and their ratios can be used to measure the AGN and starburst components in galaxies \citep{tommasin10}.}
\label{fig:fig1}
\end{figure}

Figure~\ref{fig:fig1}-a shows how well the IR fine structure lines cover the density-ionization parameter space which characterizes the photoionized and photon dissociated gas \citep[see, e.g.,][]{sm92}. 
A combination of these lines and line ratios can trace both star formation and black hole accretion. The long wavelengths of these lines, ranging from the far-IR for the photodissociation and HII region lines through the mid-IR for the AGN lines, to the near-IR for the coronal lines, ensure that we can observe these different tracers by minimizing the effect of dust extinction.

The rich rest-frame mid-IR spectra, that have been recently observed in active and starburst galaxies in the local Universe with the mid-IR spectrometer IRS \citep{houck04} onboard the Spitzer satellite \citep{Werner_2004} can be observed in the far-IR in the redshift range of 0.4$<$z$<$3.0. 

Figure~\ref{fig:fig1}-b shows the average Spitzer IRS high-resolution mid-IR spectra \citep{tommasin10} of subclasses of Seyfert galaxies from the the 12$\mu$m Seyfert galaxy sample of \citet{rms93}.  
For comparison, we also show the average spectrum of starburst galaxies \citep{b-s09}. The quality of the data is very high and shows the many features that can distinguish between AGN and star formation processes, such as the high-ionization lines from [NeV] at 14.3$\mu$m and 24.3$\mu$m originated exclusively from AGN or the 11.2$\mu$m PAH feature and the low ionization lines from [NeII] and [SIII], typical of HII and star forming regions. Mid-/far-IR imaging spectroscopy is therefore able to trace galaxy evolution throughout cosmic times in an unbiased way by minimizing dust extinction.

\begin{figure}[!ht]
\begin{center}
   \resizebox{1\hsize}{!}{
     \includegraphics*{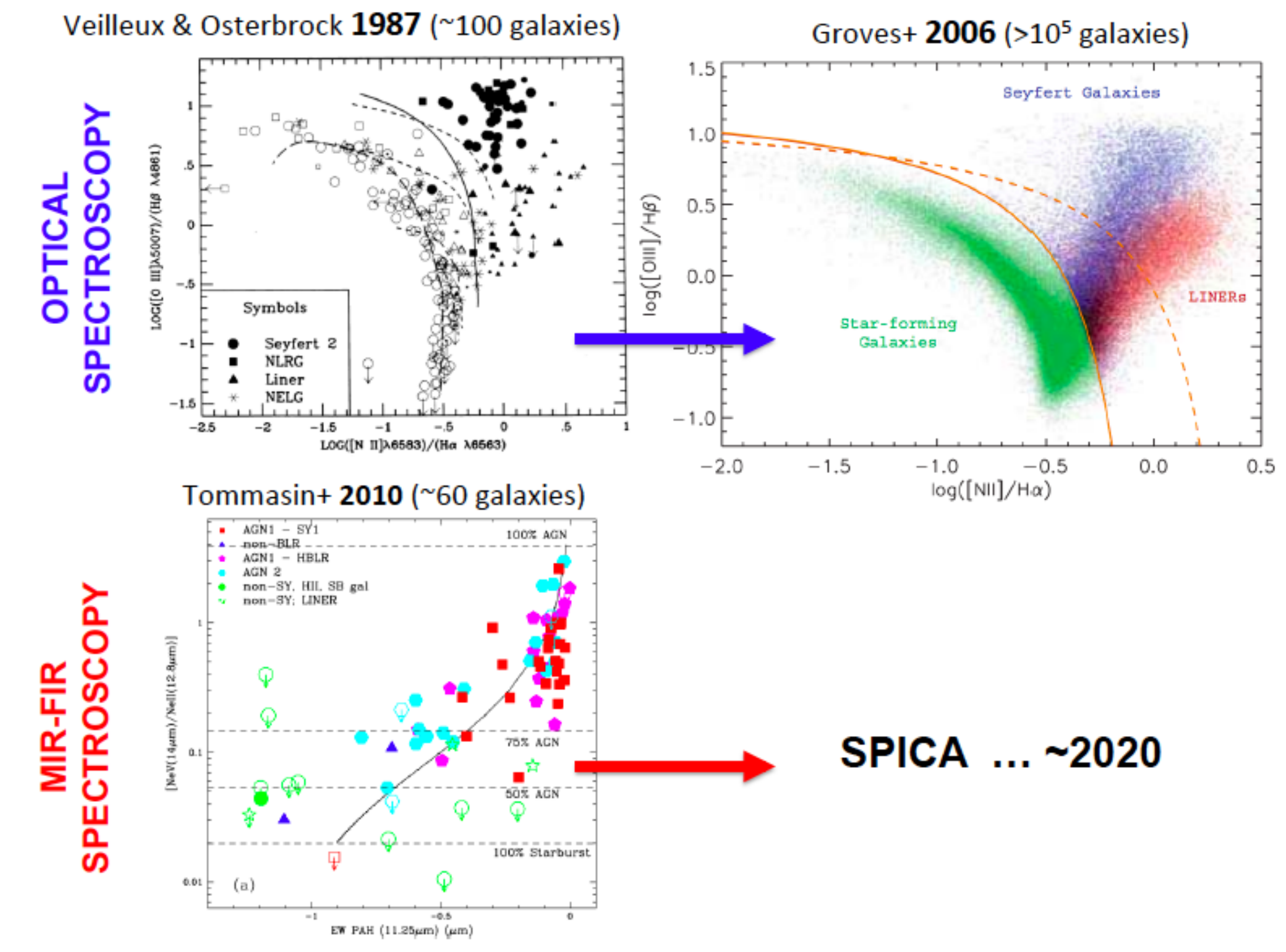}
   }
\end{center}
\caption{
An illustration of the status of extragalactic infrared spectroscopy as compared to that of optical spectroscopy: the jump from \citet{veilleux87} to \citet{groves06} has jet to be done in the infrared.
The pioneering work in active galaxies classification with infrared spectroscopy has just started with, e.g., \citet{tommasin10}.
}
\label{fig:fig2}
\end{figure}

As one can see from the line ratio diagrams in Figure~\ref{fig:fig2}, mid-IR and far-IR extragalactic spectroscopy is currently at the same stage as optical spectroscopy more than one decade ago.
Due both to the atmospheric absorption, which leaves open only a few sparse windows in the near- and mid-IR, and to
the high thermal background at room temperature at IR wavelengths, it has soon been realised that infrared astronomy 
to be successful had to be done from space telescopes, as it was demonstrated by the success of the IRAS \citep{Neugebauer_1984}, ISO \citep{Kessler_1996}, 
{\it Spitzer} \citep{Werner_2004}, AKARI \citep{Murakami_2007} and finally {\it Herschel} \citep{Pilbratt_2010} missions. 
However, due to the sensitivity limits and the poor multiplexing power of the spectrographs onboard of these spacecrafts, 
only a few limited samples of distant objects have been successfully observed \citep[e.g.][]{yan07,men09,stu10}, 
while most of the spectroscopic work has been done in the Local Universe.
Substantial progress in studying galaxy evolution therefore can only be achieved by using direct mid- to far-IR spectroscopic surveys, which will provide measured (rather than estimated) redshifts and also unambiguously characterise the detected sources, by measuring the AGN and starburst contributions to their bolometric luminosities over a wide range of cosmological epochs through the spectroscopic signatures of both AGN  and star formation emission. 

SPICA \citep{Nakagawa_2011} will be the next-generation, space infrared observatory, which, for the first time, 
will contain a large (3.2-meter) actively cooled telescope (down to 6K), providing an extremely low background 
environment. With its instrument suite, designed with state-of-the art detectors to fully exploit this low background, 
SPICA will provide not only high spatial resolution and unprecedented sensitivity in mid- and far-infrared imaging, but especially
large field medium spectral resolution imaging spectroscopy. These characteristics put SPICA among the best planned facilities to
perform spectroscopic cosmological surveys in the mid- to far-IR. Using theoretical models for galaxy formation and evolution constrained by the luminosity 
functions observed with both {\it Spitzer} and {\it Herschel} and the relations between line and continuum far-IR
luminosity, as measured in the local Universe for active and starburst galaxies, \citet{spinoglio12} have predicted, as a function of redshift, the intensities of key lines able to trace AGN
and star formation activity along cosmic history.

\begin{figure}[!ht]
\begin{center}
   \resizebox{1.\hsize}{!}{
     \includegraphics*{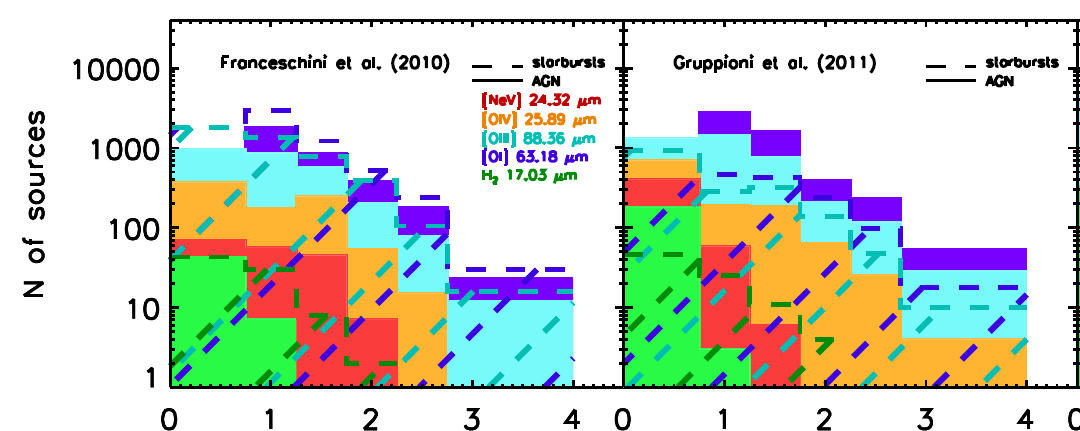}
   }
\end{center}
\caption{Number of objects detected as a function of redshift, per spectral line and per object type, (AGN-dominated are shown as continuum lines, and starburst-dominated galaxies as dashed lines) in an hour integration per pointing, 0.5 $deg^{2}$ survey with the SPICA-SAFARI instrument \citep{roel12} requiring 450 hours of total integration time \citep{spinoglio12}, following the two different models of \citet{fra10} and of \citet{gru11}.
}
\label{fig:fig3}
\end{figure}

Figure~\ref{fig:fig3} shows graphically the number of galaxies that can be detected by the far-IR FTS spectrometer SAFARI \citep{roel12} planned to be onboard of SPICA, in each spectral line for the two different populations of AGN-dominated and starburst-dominated galaxies, comparing the output of two models used. The total numbers of detectable objects agree, taking the different models, to within a factor of 2-3 for most lines and z ranges. At least a thousand galaxies will be simultaneously detected in four lines at 5$\sigma$ over a half square degree.  A survey of the given assumptions will lead to the detection of bright lines (e.g., [O I] and [O III]) and PAH features in thousands of galaxies at z$>$1. Hundreds of z$>$1 AGN will be detected in the [O IV] line, and several tens of z$>$1 sources will be detected in [Ne V] and H$_2$.

On the other hand, the Cerro Chajnantor Atacama Telescope (CCAT) \citep{seb10} will be highly complementary to SPICA, being able to observe the [OIII]88$\mu$m line at z $>$ 1.3, where this line leaves the SAFARI spectral range. We also find that CCAT will be a most efficient instrument for studies of [CII], an important coolant of the interstellar medium (ISM), at all z $<$ 5. At 3 $<$ z $<$ 4 alone, it will detect more than 300 galaxies at 5$\sigma$ level in a 0.5 deg$^2$ survey \citep{spinoglio12}.

\section{Conclusions}

We summarise this work with these points:\\
\begin{itemize}
\item[-]  After many decades of efforts, we are close to having reliable measures of star formation rate and AGN accretion power, through MIR/FIR spectroscopic surveys, unaffected by dust. 
\item[-]  Accurately measuring the star formation rate and the AGN accretion power is the first step towards understanding galaxy evolution over the history of the Universe.
\item[-] Blind FIR spectroscopic surveys with SAFARI-SPICA will be the way to ÒphysicallyÓ measure galaxy evolution.
\item[-] Given the expected sensitivity of  SAFARI-SPICA $\sim$2.5$\times$10$^{-19}$ W/m$^2$ (5$\sigma$,1 hr.) thousands of sources will be detected in more than 4 lines in typical 0.5 sq.deg. surveys (total t=450 hours).
\item[-] Complementary to SAFARI, CCAT will detect several tens to hundreds of galaxies at R$\sim$1000 in a 0.5 sq. deg. survey in 4.5 hours in the [OIII]88$\mu$m line and thousands of galaxies in the [CII]158$\mu$m line.
\item[-]These surveys will be essential to clarify the inter-relation between quasar activity and star formation, which of the two processes influence the other and ultimately will test the processes able to shape the mass and luminosity functions of galaxies.
\end{itemize}

\bibliography{lspinoglio}

\end{document}